\documentclass[twocolumn,superscriptaddress,showpacs,aps,pra,%
longbibliography]{revtex4-1}

\usepackage{epsf}
\usepackage{amsmath}
\usepackage{amsfonts}
\usepackage{amssymb}
\usepackage{bm}
\usepackage{bbm}
\usepackage{nicefrac}
\usepackage{color}
\usepackage{pifont}
\usepackage{graphicx}
\usepackage{enumerate}
\usepackage{dcolumn}
\usepackage{maybemath}

\newcommand{\maxrule}{\rule[-2mm]{0mm}{6mm}}
\newcommand{\stdrule}{\rule[-1mm]{0mm}{4mm}}

\begin{document}

\title{Magic wavelength for the hydrogen $\bm{1S}$--$\bm{2S}$ transition: \\
Contribution of the continuum and the reduced-mass correction}

\author{C. M. Adhikari}
\affiliation{Department of Physics, Missouri University of Science
and Technology, Rolla MO65409, USA}

\author{A. Kawasaki}

\affiliation{Department of Physics, MIT-Harvard Center for Ultracold Atoms and 
Research Laboratory of Electronics, \\
Massachusetts Institute of Technology, Cambridge, Massachusetts 02139, USA}

\author{U. D. Jentschura}
\affiliation{Department of Physics, Missouri University of Science
and Technology, Rolla MO65409, USA}

\begin{abstract}
Recently, we studied the magic wavelength for the atomic hydrogen
$1S$-$2S$ transition [A.K., Phys.~Rev.~A {\bf 92}, 042507 (2015)]. An explicit
summation over virtual atomic states of the discrete part of the hydrogen
spectrum was performed to evaluate the atomic polarizability.  In this
paper, we supplement the contribution of the continuum part of the spectrum
and add the reduced-mass correction.  The magic wavelength,
at which the lowest-order ac Stark shifts of the $1S$ and $2S$ states are equal,
is found to be equal to $514.6 \, {\rm nm}$. 
The ac Stark shift at the magic wavelength is 
$-221.6 \, {\rm Hz} / ({\rm kW}/{\rm cm}^2)$, 
and the slope of the ac Stark shift at the magic
wavelength under a change of the 
driving laser frequency 
is $-0.215\,7 \, {\rm Hz} / ( {\rm GHz} \, {\rm kW}/{\rm cm}^2)$.
\end{abstract}


\maketitle

%
%
\section{Introduction}

The ac Stark shift is one of the most important perturbations experienced
by atoms in external fields. It is induced by any 
oscillating electric field, and is not restricted to 
resonant driving.
On the  one hand, the presence of the ac Stark shift
is beneficial when it comes to trapping atoms by a light force in
a dispersive region. Optical dipole traps~\cite{BeWa2000} and
optical lattices are widely used in the study of 
ultracold atoms~\cite{Bl2005}. On the other hand, the 
ac Stark shift becomes an obstacle
in precision measurements aiming to determine transition frequencies
in atoms to a level of one 
Hertz or better. The frequency of narrow two-photon
transitions induced by an intense light field,
is shifted significantly by the exciting
light field~\cite{HaEtAl2006,PaEtAl2011}.
The ac Stark shift due to
the black-body radiation is one of the major uncertainties in optical
lattice-clock experiments~\cite{UsEtAl2016}, and the ac Stark shift due to
the optical lattice light is an essential effect in optical lattice
clock experiments~\cite{NiEtAl2015,KaTaPaOv2003}. 

When the optical dipole trap or the optical lattice clock is used for trapping
atoms, the resulting ac Stark shift is significantly larger than the 
target precision for the transition frequency, and furthermore,
the ac Stark shift is
generally different for the ground state and the excited state of the
transition. To cancel this shift, one needs to use the light of 
a special wavelength called the magic wavelength~\cite{KaTaPaOv2003}. 

%
%
\section{Calculation of the polarizability}

In order to calculate the magic wavelength, we first evaluate the ac Stark shift
for the ground state and the excited state of a transition, and then search the
point where the difference is zero. The ac Stark is given for a specific atomic
reference state $|\phi \rangle$ as 
\begin{equation}
\delta E_{\rm ac} = - \frac{I_L}{2 \epsilon_0 c} \, \alpha(\phi, \omega_L) \,,
\end{equation}
where $I_L$ is the laser intensity, and $\alpha(\phi, \omega_L)$, $| \phi
\rangle$, and $\omega_L$ are the dipole polarizability, atomic reference
state, and angular frequency of the laser,
respectively~\cite{HaJeKe2006,JePa2015epjd1}.

The dipole polarizability for a reference state $| \phi \rangle$, whose energy
is denoted as $E$, reads as
\begin{align} 
\label{form}
\alpha(\phi, \omega_L) =& \;
\frac{e^2}{3} \, \sum_\pm
\left< \phi \left| \vec r \,
\left( \frac{1}{H_A - E \pm \hbar \omega_L} \right) \, \vec r
\right| \phi \right> 
\nonumber\\[0.1133ex]
=& \; P_\phi(\omega_L) + P_\phi(-\omega_L) \,,
\end{align}
where $\vec r$ is the electron position 
operator (the scalar product is implied by the 
repeated occurrence of the vector). 
Furthermore, $H_A$ is the atomic (Schr\"{o}dinger) Hamiltonian.
The $P$ matrix elements are implicitly defined as
the terms that emerge from the sum over virtual states.

We have already discussed~\cite{Ka2015} that 
optical trapping with light of the magic wavelength 
corresponding to the hydrogen $1S$-$2S$
transition could be important for improving Doppler-free two-photon
spectroscopy.  The calculation described in Ref.~\cite{Ka2015} 
included the contribution from the discrete virtual states.
With the formalism of Eq.~\eqref{form}, we include the effect of the continuous
part of the spectrum as well. 
Based on Ref.~\cite{CaPeSe1960},
we know that the latter effect can be large for 
the dc Stark shift of the hydrogen ground state.
It is known that matrix elements of the
form~\eqref{form} can be summed in close analytic 
form~\cite{GaCo1970,SwDr1991a,SwDr1991b,SwDr1991c,Pa1993,JePa1996}.  Angular
components are calculated separately from the radial
components~\cite{Ed1974ang}.  As for the radial component, one first performs
the Sturmian decomposition of the Schr\"{o}dinger Green 
function~\cite{Je1996}, then does the radial integrations, and 
finally carries out the
summations over the discrete and continuous spectra.

In the approximation of an infinite nuclear mass, the calculation for the
hydrogen $1S$ state results in
\begin{subequations}
\begin{align}
\label{pol1S}
& P_{1S}(\omega_L) = -
\frac{e^2 \, a_0^2}{E_h} \,
\left[ \frac{2t^2}{3 (1-t)^5 (1 + t)^4}
\left(38 t^7
+ 26 t^6 \right.\right.\nonumber\\[0.133ex]
&\left.\left. + 19 t^5 - 19 t^4 - 12 t^3 + 12 t^2 + 3 t - 3\right)
\right.
\nonumber\\[0.133ex]
& \left. -
\frac{256 \, t^9}{3 \, (t - 1)^5 \, (t + 1)^5} \,
{}_2 F_1\left(1, -t, 1 - t, \left( \frac{1-t}{1+t} \right)^2 \right) \right],
\nonumber\\[0.133ex]
& t = \left( 1 + \frac{2 \hbar \omega_L}{E_h} \right)^{-1/2} \,,
\end{align}
whereas one obtains for $2S$,
\begin{align}
\label{pol2S}
& P_{2S}(\omega_L) =
\frac{e^2 \, a_0^2}{E_h} \,
\bigg[ \frac{16\tau^2}{3 (\tau-1)^6 (1 + \tau)^4}
(1181 \tau^8 - 314 \tau^7
\nonumber\\[0.133ex]
&
- 16 \tau^6 - 166 \tau^5
+ 14 \tau^4 + 138 \tau^3 - 48 \tau^2 - 42 \tau + 21)
\nonumber\\[0.133ex]
& -
\frac{16384 \, \tau^9 \, (4 \tau^2 - 1)}{3 \, (\tau - 1)^6 \, (\tau + 1)^6} \,
{}_2 F_1\left(1, -2 \tau, 1 - 2 \tau, \left( \frac{1- \tau}{1+\tau} \right)^2
\right) \bigg] \,,
\nonumber\\[0.133ex]
& \tau = \left( 1 + \frac{8 \hbar \omega_L}{E_h} \right)^{-1/2} \,.
\end{align}
\end{subequations}
Here, $a_0 = \hbar/(\alpha m_e c)$, $m_e$, and $E_h = \alpha^2 m_e c^2$ are the
Bohr radius, the electron mass, and the Hartree energy, respectively.
The complete (Gaussian) hypergeometric function is 
denoted as ${}_2 F_1$.

The magic angular frequency $\omega_M$ is determined 
by the condition $f(\omega_L = \omega_M) = 0$, where 
\begin{equation}
f(\omega_L) = \alpha(2S, \omega_L) -
\alpha(1S, \omega_L) \,.
\end{equation}
An evaluation using the Newton-Raphson technique with a starting value of
$\hbar \omega_M \approx 0.09 \, E_h$, which is a one-significant-digit 
approximation of the magic wavelength inspired by 
our previous calculation~\cite{Ka2015}, converges
to a value of $\hbar \omega_M \approx 0.088\,581\,526 \, E_h$.
We employ
quadruple precision arithmetic (32 decimals) in intermediate steps. 

The first reduced-mass correction is taken into 
account by observing that the hydrogen transition frequencies,
and transition matrix elements, scale with the 
reduced mass of atomic hydrogen, 
\begin{equation}
m_r = \frac{m_e \, m_p}{m_e + m_p} \,,
\end{equation}
where $m_p$ is the proton mass.
The result for the magic angular frequency thus receives 
an additional correction factor $m_r/m_e$ and reads as
\begin{equation}
\omega_M = 2 \pi \times 5.825\,211 \times 10^{14} \, {\rm Hz} \,,
\end{equation}
which corresponds to a frequency of 
$\nu_M = 5.825\,211 \times 10^{14} \, {\rm Hz}$.
The magic wavelength thus is
\begin{equation}
\label{lambdaMAGIC}
\lambda_M = 514.646 \, {\rm nm} \,.
\end{equation}

The difference between the result for the magic wavelength 
obtained here [Eq.~\eqref{lambdaMAGIC}] 
and our previous calculation~\cite{Ka2015} 
is larger than the reduced-mass correction. 
This implies that the effect of the continuous part
of the spectrum is not negligible.
This is consistent with observations made
in the calculation of the dc
Stark shift of the hydrogen ground state,
and Bethe logarithms in other simple
atomic systems like helium~\cite{DrGo1999}.  As evident from Fig.~1 of
Ref.~\cite{Ka2015}, the wavelength~\eqref{lambdaMAGIC}
lies in between the $2S$--$3S$ and
$2S$--$4S$ transitions. We note that the vertical bars in the cited figure
correspond to the sign changes of the ac Stark shift near resonant frequencies
of the hydrogen atom; these resonances formally induce poles as they correspond
to zeros of the propagator denominator in Eq.~\eqref{form}.

To evaluate the absolute value of the ac Stark shift at the magic
wavelength numerically, the series representation~\cite{Ba1953vol1,Ba1953vol2}
of the hypergeometric function is sufficient. The result is
\begin{align}
\label{EacMAGIC}
\Delta E_M = & \; \Delta E_{\rm ac}(1S, \omega_M) = 
\Delta E_{\rm ac}(2S, \omega_M) 
\nonumber\\[0.1133ex]
=& \; -221.584 \, \frac{I_L}{{\rm kW}/{\rm cm}^2} \, {\rm Hz} \,.
\end{align}
In obtaining Eq.~\eqref{EacMAGIC}, we have 
taken into account that the polarizability matrix 
elements in Eqs.~\eqref{pol1S} and~\eqref{pol2S} receive
reduced-mass corrections in the form of  
factors $(m_e/m_r)$, which multiply the 
Bohr radius and the energy denominator, resulting in an
overall prefactor $(m_e/m_r)^3$.
Finally, the slope of the ac Stark shift within the improved 
formulation of the problem presented in this paper is 
\begin{align}
\label{slopeMAGIC}
\eta =& \; \left. \frac{\partial}{\partial \omega_L} 
\big( \Delta E_{\rm ac}(2S, \omega_M) -
\Delta E_{\rm ac}(1S, \omega_M) \big) \right|_{\omega_L = \omega_M} 
\nonumber\\[0.1133ex]
=& \; -0.215\,748 \, \frac{{\rm Hz}}{{\rm GHz} \, ({\rm kW}/{\rm cm}^2)} \, .
\end{align}

%
%
\section{Relativistic and field-configuration corrections}

Relativistic corrections to the polarizability can be taken into account, if
desired, by perturbing the Hamiltonian, wave function, and the
energy of the reference state, in the following way,
\begin{subequations}
\begin{align}
H_A \to & \; H_A + H_R \,,
\\[0.1133ex]
E \to & \; E + \langle H_R \rangle \,,
\\[0.1133ex]
|\phi \rangle  \to & \; |\phi \rangle + 
\left( \frac{1}{ E - H_A} \right)' \, H_R \, |\phi\rangle \,.
\end{align}
\end{subequations}
Here, the Schr\"{o}dinger Hamiltonian $H_A$ and the relativistic correction term $H_R$ are
\begin{eqnarray}
H_A &= & \frac{\vec{p}^{\,2}}{2 m_e} - \frac{\alpha \, \hbar \, c}{r}  \label{HR} \\
H_R &= & \; -\frac{\vec{p}^{\,4}}{8 m_e^3 c^2}
+ \frac{1}{2} \alpha\left(\frac{\hbar^2 g_s}{2m^{2} \, c}\right)
\frac{\vec{L}\cdot\vec{S}}{| \vec r |^3}
\nonumber\\
& & + \frac{\hbar^{3}}{8 m_e^2 c}\,4\pi\alpha \,
\delta^{(3)}(\vec{r})  \,,
\end{eqnarray}
where $g_s \approx 2$ is the spin $g$ factor.
These relativistic effects shift the transition frequencies in hydrogen,
and the magic wavelength, by a relative correction of order $\alpha^2 \sim 10^{-4}$.
The relative accuracy of the results given in 
Eqs.~\eqref{lambdaMAGIC},~\eqref{EacMAGIC} and~\eqref{slopeMAGIC}
thus is of the order of $10^{-4}$.
The reduced-mass correction, by contrast, is of the 
order of $m_e/m_p \sim 10^{-3}$ and is the dominant
correction to the nonrelativistic one-particle approximation.

\begin{table}[th!]
\caption{\label{table1} Influence of the reduced-mass 
correction (RMC) on the magic wavelength $\lambda_M$,
ac Stark shift $\Delta E_M$, and
slope $\eta$ of the ac Stark shift,
with results indicated in Eqs.~\eqref{lambdaMAGIC},~\eqref{EacMAGIC},
and~\eqref{slopeMAGIC}. The $1S$--$3S$ and $1S$--$4S$ results
are obtained using a generalization of the result given in 
Eq.~\eqref{pol2S} to higher excited states, using techniques 
familiar from analytic Lamb shift calculations~\cite{JeCzPa2005}.}
\begin{center}
\begin{tabular}{ccc}
\hline
\hline
\multicolumn{1}{c}{Quantity} & 
\multicolumn{1}{c}{Without RMC} &
\multicolumn{1}{c}{With RMC} \\
\hline
\multicolumn{3}{c}{\maxrule $1S$--$2S$ Transition} \\
\stdrule
$\lambda_M$ &
$514.366 \,{\rm nm}$ & 
$514.646\, {\rm nm}$ \\
\stdrule
$\Delta E_M$ &
$-221.222 \, \frac{I_L}{{\rm kW}/{\rm cm}^2} \, {\rm Hz}$ & 
$-221.584 \, \frac{I_L}{{\rm kW}/{\rm cm}^2} \, {\rm Hz}$ \\
\stdrule
$\eta$ &
$-0.215\,396 \, \frac{{\rm Hz}}{{\rm GHz} \, ({\rm kW}/{\rm cm}^2)}$ & 
$-0.215\,748 \, \frac{{\rm Hz}}{{\rm GHz} \, ({\rm kW}/{\rm cm}^2)}$ \\
\multicolumn{3}{c}{\maxrule $1S$--$3S$ Transition} \\
\stdrule
$\lambda_M$ &
$1371.11 \,{\rm nm}$ &
$1371.85 \, {\rm nm}$ \\
\stdrule
$\Delta E_M$ &
$-212.307 \, \frac{I_L}{{\rm kW}/{\rm cm}^2} \, {\rm Hz}$ &
$-212.654 \, \frac{I_L}{{\rm kW}/{\rm cm}^2} \, {\rm Hz}$ \\
\stdrule
$\eta$ &
$-3.206\,79 \, \frac{{\rm Hz}}{{\rm GHz} \, ({\rm kW}/{\rm cm}^2)}$ &
$-3.212\,03 \, \frac{{\rm Hz}}{{\rm GHz} \, ({\rm kW}/{\rm cm}^2)}$ \\
\multicolumn{3}{c}{\maxrule $1S$--$4S$ Transition} \\
\stdrule
$\lambda_M$ &
$2811.24 \,{\rm nm}$ &
$2812.77\, {\rm nm}$ \\
\stdrule
$\Delta E_M$ &
$-211.249 \, \frac{I_L}{{\rm kW}/{\rm cm}^2} \, {\rm Hz}$ &
$-211.594 \, \frac{I_L}{{\rm kW}/{\rm cm}^2} \, {\rm Hz}$ \\
\stdrule
$\eta$ &
$-28.467\,6 \, \frac{{\rm Hz}}{{\rm GHz} \, ({\rm kW}/{\rm cm}^2)}$ &
$-28.514\,2 \, \frac{{\rm Hz}}{{\rm GHz} \, ({\rm kW}/{\rm cm}^2)}$ \\
\hline
\hline
\end{tabular}
\end{center}
\end{table}

There is, in addition, a field-configuration dependent shift of the 
transition frequency, due to the following term in 
the long-wavelength quantum electrodynamic Hamiltonian \cite{Pa1993,EvJeKe2004}.
\begin{align}
\label{LWpotential}
H_{LW} =& -e\,\vec r \cdot \vec E(t,\vec {0})
-\frac{e}{2}\,r^i\,r^j\,
\left. \frac{\partial E^i(t,\vec{r})}{\partial r^j}\right|_{\vec{r}=\vec{0}}
\nonumber\\
&-\frac{e}{6}\,r^i\,r^j\,r^k\,
\left. \frac{\partial^2 E^i(t,\vec{r})}{\partial r^j \partial r^k}
\right|_{\vec{r}=\vec{0}}\,.
\end{align}
Let us assume,
for definiteness, a plane standing wave of linearly $z$-polarized light with wave vector
$\vec k$ aligned along the $x$-direction. In this case,
the electric field is given by 
\begin{equation}
\label{laserconf}
\vec{E}(t, x) = \hat{e}_z \, {\cal E}_L \, \cos(\omega_L t) \, \cos(k_L x)\,,
\end{equation}
where $k_L = \omega_L/c$ and ${\cal E}_L$ is the peak electric field
during a laser cycle.
We assume that atoms are at antinodes of the standing wave, 
i.e., that we have $\cos(k_L \, x) = 1$ at the position of the atom.
In this case, the first and third terms in~\eqref{LWpotential}
contribute, and we obtain
\begin{eqnarray}
\label{VLW}
H_{LW} \approx -e \, z \, {\cal E}_L \cos(\omega_L t) +
\frac{e}{6} k_L^2 \, x^2 \, z \, {\cal E}_L \cos(\omega_L t) \,.
\end{eqnarray}
The leading field-configuration 
dependent correction to the dynamic polarizability of state
$|\phi\rangle$ therefore reads
\begin{equation}
\label{FieldCorrLeading}
\delta \alpha(\phi, \omega_L) =
-\frac{e^2 \, k_L^2}{6}\sum\limits_\pm\displaystyle \left< \phi \left|\,
z\,\frac{1}{H_0-E_\phi\pm\hbar\omega_L}\,x^2 z\, \right| \phi \right>\,,
\end{equation}
but this expression depends on our choice~\eqref{laserconf} of the laser 
field configuration and would be different for, e.g.,
a traveling as opposed to standing wave.
The second term on the right-hand side
of Eq.~\eqref{LWpotential}, which is a lower-order contribution, 
vanishes for symmetry reasons,
and magnetic effects can be neglected~\cite{Ka2015}.
As already stated, the magic angular frequency for the two-photon
$1S$--$2S$ transition lies in between the frequencies 
of the single-photon $2S$--$3P$ and $2S$--$4P$ transitions,
and therefore is of the same order-of-magnitude as typical
optical transition frequencies; thus, we have 
$(k_L \, x) \sim {\mathcal O}(\alpha^2)$ as a parametric estimate.
The correction~\eqref{FieldCorrLeading}
therefore is of the same order-of-magnitude as the 
relativistic correction induced by the Hamiltonian~\eqref{HR}.
Because the former depends on the specific configuration of the 
light field used in the experiment,
we do not pursue the calculation of these effects any further 
here. If needed, they can be evaluated based on techniques 
used in Lamb shift calculations~\cite{CzJePa2005}. 

%
%
\section{Conclusions}

The analysis presented in this paper will be important for any future
experimental implementation of the proposal presented in Ref.~\cite{Ka2015}.
The main results for the $1S$--$2S$ transition
are summarized as the magic wavelength at $514.646\,{\rm nm}$,
the polarizability of $-221.584 \, \frac{I_L}{{\rm kW}/{\rm cm}^2} \, {\rm Hz}$,
and its slope of 
$-0.215\,748 \, {\rm Hz}/({\rm GHz} \, ({\rm kW}/{\rm cm}^2))$. 
These results are separately indicated in Table~\ref{table1},
with a focus on the reduced-mass correction,
and results for the magic wavelengths and ac Stark effects of the 
$1S$--$3S$ and $1S$--$4S$ transitions are supplemented (cf.~Ref.~\cite{YiZhLiZh2016}).
The theoretical uncertainty of these values is on the level of
$10^{-4}$. Nevertheless, we
have adopted the policy of indicating the numerical results to a nominal accuracy of six
decimals, in order to facilitate an independent numerical evaluation.
The dominant correction to the nonrelativistic one-particle approximation 
of the magic wavelength
is due to the reduced-mass correction, and the relativistic 
correction of order $\alpha^2$ is shadowed by a laser-field configuration
dependent correction which has to be individually evaluated 
for a particular experimental setup.

%
%
\section*{Acknowlegemnts}

This research was supported by the National Science Foundation
under Grant PHY--1403973.

\end{document}